\documentclass[usenatbib]{mn2e}
\usepackage{graphicx}

\newcommand{\ltaraw}{$\; \buildrel < \over \sim \;$}
\newcommand{\lta}{\lower.5ex\hbox{\ltaraw}}
\newcommand{\gtaraw}{$\; \buildrel > \over \sim \;$}
\newcommand{\gta}{\lower.5ex\hbox{\gtaraw}}

\loadboldmathitalic   \title[Not the Cosmic Horizon]
{ Through the Looking Glass: 
  Why the ``Cosmic Horizon'' is not a horizon\footnotemark[1]
} 
 \author[van Oirschot et al.]
{ Pim van Oirschot$^{1,2}\footnotemark[2]$,
  Juliana Kwan$^1$ \& Geraint F. Lewis$^{1}$\\
  $^{1}$Sydney Institute for Astronomy, School of Physics, A29,
  University of Sydney, NSW 2006, Australia\\
  $^{2}$Department of Astrophysics, IMAPP, 
  Radboud University Nijmegen, PO Box 9010, 
  6500 GL Nijmegen, The Netherlands\\
}
\date{\today}
\begin{document}
\maketitle
\label{firstpage}
\begin{abstract}
The present standard model of cosmology, $\Lambda$CDM, contains some
intriguing coincidences. Not only are the dominant contributions to
the energy density approximately of the same order at the present
epoch, but we note that contrary to the emergence of cosmic
acceleration as a recent phenomenon, the time averaged value of the
deceleration parameter over the age of the universe is nearly
zero. Curious features like these in $\Lambda$CDM give rise to a
number of alternate cosmologies being proposed to remove them,
including models with an equation of state $w = -1/3$. 
In this paper, we examine the validity of some of these 
alternate models and we also address some persistent 
misconceptions about the Hubble sphere and the event
horizon that lead to erroneous conclusions about cosmology.

 \end{abstract}

\long\def\symbolfootnote[#1]#2{\begingroup%
  \def\thefootnote{\fnsymbol{footnote}}\footnotetext[#1]{#2}\endgroup} 

\def\newblock{\hskip .11em plus .33em minus .07em}
\section{Introduction}     \label{introduction}     \symbolfootnote[1]
{ Research undertaken as part of the Commonwealth Cosmology Initiative  
  (CCI: www.thecci.org), an international collaboration supported  
  by the Australian Research Council
}
\symbolfootnote[2]
{ pimvanoirschot@gmail.com
}

\begin{figure*} 
\begin{center}
\includegraphics[scale = 0.5,angle=270]{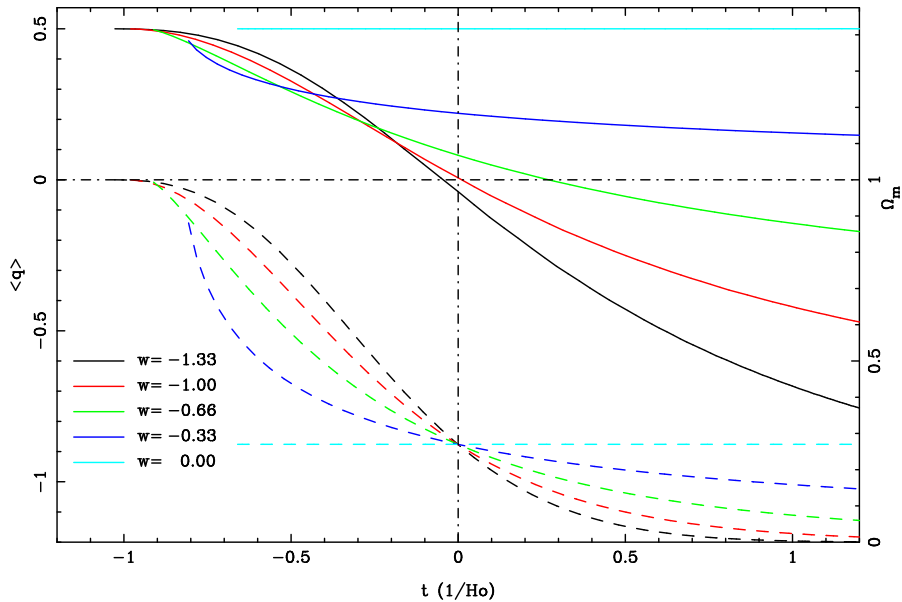}
\caption{The solid lines represent the time averaged value of 
the deceleration parameter $\langle q \rangle$ for flat cosmologies with 
a density parameter of dark energy, $\Omega_{\mathrm{X},0} = 0.726$ 
and different equations of state, $w$, for the dark energy component.
The dashed lines represent the evolution of the density parameter of
matter, $\Omega_\mathrm{m}$, in the same models.
The dot-dashed line at $t=0$ corresponds to the present epoch,
while the dot-dashed line at $\langle q \rangle = 0$ corresponds
to a time averaged deceleration of 0. Note that only the red solid
line (corresponding to $\Lambda$CDM) goes through 
the intersection of these two lines.}
\label{coincidence}
\end{center}
\end{figure*} 

There is growing observational evidence for the existence of a
non-zero cosmological constant \citep{Riess,
Perlmutter,Spergel,Tegmark}, yet there many alternative theories for
cosmic acceleration as a number of outstanding, fundamental questions
concerning the $\Lambda$ Cold Dark Matter ($\Lambda$CDM) paradigm
remain unsolved. A key problem with the cosmological constant is that
its energy density derived from observations, $\Omega_{\Lambda},$ is
$\approx$120 orders of magnitude smaller than what we would expect from the
predictions of quantum field theory~\citep{Weinberg}. Also, it is
sometimes remarked that the near equality between the best fitting values
of $\Omega_{\Lambda}$ and $\Omega_{m}$ obtained for $\Lambda$CDM
presents a ``coincidence problem'', since it implies that we are placed
at a special time in cosmic history when the energy densities are
approximately equal. There have been numerous attempts to remedy these
problems, such as evolving dark energy [for a review, see
\citet{Barnes}], but none of these are particularly convincing or well
supported by observations and $\Lambda$CDM remains the standard model
of cosmology.

One recent alternative model was dependent upon the properties of the
 ``cosmic horizon'', $R_\mathrm{h}$, defined by \citet{melia2007} 
 as a Schwarzschild radius $R_\mathrm{h} = 2GM(R_\mathrm{h})$ 
 (throughout this paper, the speed of light is set equal to unity).
In a Friedmann-Lema\^itre-Robbertson-Walker (FLRW)
universe, with flat spatial geometry, $R_\mathrm{h}$
is equal to $1/H$, where $H(t)$ is the Hubble parameter.
\citet{melia2009} showed that the time derivative
(denoted by an overdot) of the ``cosmic horizon''
\begin{equation}
\dot{R}_\mathrm{h} = \frac{3(1+w)}{2}, \label{Rhdot}
\end{equation}
for a single component universe in which
the cosmic fluid has an equation of state $w$.
Note that $\dot{R}_\mathrm{h} = 1$ only for the
special case of $w = -1/3$, thus $R_\mathrm{h}$
is exactly equal to $t$ at all times in such a universe.

From the present day best fitting value
$\Omega_{\Lambda,0} = 0.726\pm0.015$~\citep{komatsu} (a subscript 
zero denotes the value of a quantity at the present time)
and assuming a spatially flat universe, 
it can be derived that our universe is approximately
13.7 billion years old. Using the value
$H_0 = 70.5 \pm 1.3 \ \mathrm{km} \ \mathrm{s}^{-1} \ \mathrm{Mpc}^{-1}$~\citep{komatsu},
this age can be written as 0.989 Hubble time ($1/H_0$), thus 
\begin{equation}
t_0 \approx \frac{1}{H_0} = R_\mathrm{h}(t_0).\label{coincidence2}
\end{equation}

\citet{melia2009} and Melia \& Abdelqader (2009; hereafter MA09)
argues that this equality (or near equality)
should signify that the best cosmological model is one in
which these quantities are equal for all cosmic times, 
i.e. the $w = -1/3$ model mentioned above, not just for a
brief crossing that happens to occur now. 
However, as we shall argue in Section~\ref{problem},
this model relies the ability of `anthropic' reasoning 
to discriminate between cosmologies.
Furthermore, the model proposed by~MA09 requires
$R_\mathrm{h}$ to act as a true horizon, such that the
redshifting of photons emanating from this surface becomes infinite.
We shall show in Section~\ref{horizons} and~\ref{redshifting} 
that this assertion is erroneous and that the conclusions 
presented in~MA09 rely on a misapplication of the Hubble
sphere. The goal of this paper is to clarify some of the pernicious 
misconceptions surrounding the Hubble sphere and to address 
the validity of the ``cosmic horizon'' as a test of cosmology.

\section{Curiouser and curiouser: One coincidence problem becomes two.}\label{problem}

MA09 argues that since $R_\mathrm{h}$ would equal $t$ 
just once in the entire history of the universe if $w \neq -1/3$,
it is an unacceptably improbable coincidence that $R_\mathrm{h}\approx t_0$ 
right now. In this section, we shall discuss the near equality of 
$R_\mathrm{h}$ and $t_0$ and show that it indeed poses an additional 
coincidence problem for $\Lambda$CDM.
However, we argue that Equation~\ref{coincidence2} can not be used as the
basis for constructing a cosmological model that is competitive with
$\Lambda$CDM.
Furthermore, instead of expressing the coincidence in terms of $R_\mathrm{h}$,
we shall express it in terms of the the average value of the
deceleration parameter $q(t)$ over the age of the universe, $\langle q(t_0) \rangle$.

The deceleration parameter is defined in terms of the scale factor
$a(t)$, which embodies the evolutionary path of the universe, 
and it can be shown that for a flat FLRW universe [see for example \citet{Barnes}]
\begin{equation}
q \equiv - \frac{\ddot{a}a}{\dot{a}^2} = \frac{1+3w}{2}. \label{q}
\label{decel}
\end{equation}
Comparing Equations~\ref{Rhdot}
and~\ref{q}, we see that $\dot{R}_\mathrm{h} = q + 1$.
This yields that the time averaged deceleration parameter
\begin{equation}
\langle q(t) \rangle = \frac{1}{t}\int_0^{t} \left(\dot{R}_\mathrm{h}(t') -1 \right) \mathrm{d}t' = \frac{1}{t H} - 1. 
\end{equation}
Inserting the above mentioned values for $t_0$ and $H_0$, this expression gives 
$\langle q(t_0) \rangle = 0.0113\pm0.0154$ and the present average deceleration of the
universe is remarkably close to zero. 
We shall assign the fact that $\langle q(t_0) \rangle$ is consistent with zero
as a coincidence, but we note that it is a separate coincidence from the well
known ``coincidence problem'' and in fact the duration of this event
in cosmic history is so brief that it is a ``greater'' coincidence in
this respect than the approximate equality of the dominant energy
densities.

Figure~\ref{coincidence} shows both coincidences for 
a flat FLRW universe with matter and dark energy.
We use the present day value of $\Omega_{\mathrm{X},0} = 0.726$
from~\citet{komatsu} for the density parameter of dark energy,
and assume that the universe is spatially flat.
The evolution of $\langle q \rangle$, visualised by the solid lines,
can be read on the left axis, while the change of $\Omega_{\mathrm{m}}$ over time,
visualised by the dashed lines, can be read on the right axis.
The colours represent different values of the equation of state of dark energy, $w$.
Red corresponds to $w = -1$, which is true for the cosmological constant.
While the red dashed line drops from 1 to 0 in about two Hubble times,
the red solid line indicates that $\langle q \rangle \approx 0$ 
only for a fraction of a Hubble time.
Thus, the fact that the average value of the
deceleration parameter over the age of the universe is nearly
zero for $\Lambda$CDM, really is a ``greater'' coincidence
then the well known ``coincidence problem''.

Of course, it can be argued that perhaps we do not 
reside in the concordance cosmology and that this
actually signifies a failing of the standard model. In fact, this
``new'' coincidence was previously noticed by~\citet{Lima}, who thereafter
suggested that the universe evolves through a cascade of alternately
accelerating and deceleration regimes. But the origin of the physical
mechanism responsible for these oscillations remains unknown, such
that this model raises more questions than it answers. Similarly, in
response to the coincidence that $R_\mathrm{h} \approx t_0$,~MA09 
proposes that a model containing only a single fluid with $w =
-1/3$ is a better fit to the observational data, since this would give
rise to a ``cosmic horizon'' that is fixed for all time. 
But, as soon as we include matter in our cosmology, 
$R_\mathrm{h}$ approaches $t_0$ only in the infinite future,
and the fact that we observe the near equality of $R_\mathrm{h}$ and $t_0$
today suggests that the equation of state of dark energy is
probably not $-1/3$ (the blue solid line in Figure~\ref{coincidence}
clearly does not cross $\langle q \rangle = 0$).

With extent observational data, we can already provide robust
constraints on the equation of state parameter of dark energy, which currently imply
a value of $w = -1.12 \pm 0.12$~\citep{shoes}. For a model to be
competitive with the standard model, it is not sufficient to remove a
single outstanding problem with $\Lambda$CDM but it must also satisfy
the areas that $\Lambda$CDM does model well. Setting aside these
objections, in the following sections, we investigate the cosmological
model proposed by~MA09 to solve the coincidence problem by
focusing on the conceptual arguments that underpin the model instead.

\section{The Cosmological Framework}

The application of the cosmological principle of perfect homogeneity
and isotropy uniquely determines the spacetime geometry of the
standard cosmological model which is most simply encapsulated by the
FLRW metric as follows:
\begin{equation}
\mathrm{d}s^2 = \mathrm{d}t^2 - a^2(t)\left[ \frac{\mathrm{d}r^2}{1-kr^2} 
+ r^2(\mathrm{d}\theta^2 + \mathrm{sin}^2\theta\ \mathrm{d}\phi^2) \right].
\label{flrw}
\end{equation}
where $t$ represents the cosmic time (the time measured by an
observer that is spatially stationary in the above coordinates)
and ($r,\theta,\phi$) are spherical comoving coordinates.
The curvature parameter $k$ is $+1$ for a closed universe,
0 for a flat universe or $-1$ for an open universe.

This metric may be written in a number of different, 
but equivalent forms via a coordinate transformation for convenience. 
In our discussion, it is most expedient to use conformal and the 
observer-dependent coordinates of MA09, while restricting
our attention to a flat universe with two dimensions $(t,r)$.
Note that the discussion could be trivially extended to 
include all four dimensions but this does not affect the 
main thrust of our arguments. 

After applying the transformation $d\eta = dt/a(t)$ to Equation~\ref{flrw},
the conformal form of the FLRW metric reads
\begin{equation}
\mathrm{d}s^2 = a^2(\eta) 
\left[\mathrm{d} \eta^2 - \mathrm{d} r^2 \right], \label{conformalmetric}
\end{equation}
The time coordinate is now given by $\eta$, but it does not correspond to any observer\footnote{
Often, the conformal radial coordinate is denoted by $\chi$, but since
we consider a flat universe, we can keep the symbol $r$.}. 
Since photons travel along null geodesics ($\mathrm{d}s = 0$) 
in the radial direction, we find from Equation~\ref{conformalmetric} 
that lightcones in conformal coordinates are determined by
\begin{equation}
\mathrm{d}r = \pm \mathrm{d}\eta;
\label{light}
\end{equation}
thus light rays follow straight lines at $\pm$45\degr~angles when the
metric is conformal, which makes them useful for making causal
comparisons, such as those implied by cosmic horizons.

The observer-dependent form of Equation~\ref{flrw}, as derived by~MA09, is
given by
\begin{equation}
\mathrm{d}s^2 = \Phi \left[ \mathrm{d}t + \left( \frac{R}{R_\mathrm{h}} \right) 
\Phi^{-1} \mathrm{d}R \right]^2 - \Phi^{-1} \mathrm{d}R^2, \label{meliametric}
\end{equation}
where $\Phi \equiv 1 - \big( \frac{R}{R_\mathrm{h}} \big)^2$
and the radial coordinate $R(t)$ is related to the comoving
distance $r$ via
\begin{equation}
R \equiv a(t)\ r, \label{Rar}
\end{equation}
that is, $R$ is equivalent to the proper distance and a comoving
observer does not remain stationary with respect to the spatial coordinates of
this metric.
The significance of the term $R_\mathrm{h} $ will be addressed in the 
following sections, but for now it is sufficient to note that 
a singularity occurs in the metric when $R \rightarrow R_\mathrm{h}$.

\begin{figure}
\begin{center}
\input{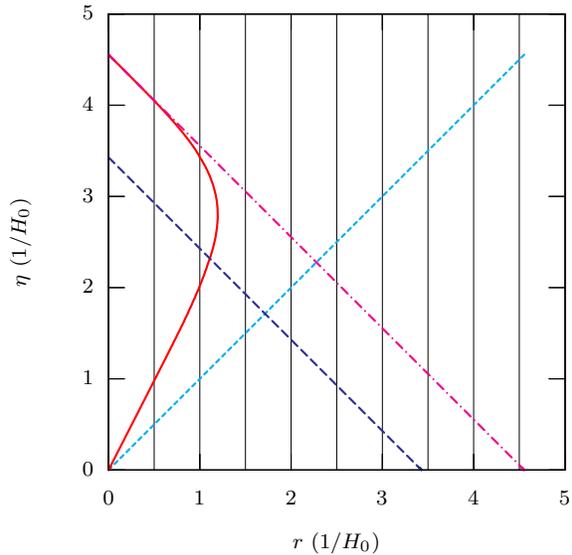}
\caption{Spacetime diagram in conformal coordinates to illustrate the
event horizon (magenta), particle horizon (cyan), Hubble surface
(red) and the past lightcone (blue). The thin black lines
illustrate the paths of comoving observers. The cosmological
parameters used were $\Omega_{\Lambda,0} = 0.726$ and $\Omega_{m,0} =
0.274$. Clearly, the Hubble sphere never coincides with the event
horizon, rather it asymptotically approaches it as $\eta \rightarrow
\eta_\mathrm{max}$. [See also, Figure~2 in \citet{Gud}, Figure~1 in
\citet{Davis}, or Figure~12.2 in \citet{Longair}.]}
\label{conformaldiagram}
\end{center}
\end{figure}

\section{Horizons in Cosmology}\label{horizons}

There are three main features when considering cosmological spacetime
diagrams in general: the event horizon, the particle horizon and the
Hubble sphere. The event horizon is defined by the surface in spacetime 
that encloses all events that can ever be detected for a comoving
observer at $t \rightarrow \infty$, that is, it consists of a lightcone 
projected backwards at the end of conformal time (see the magenta curve
in Figure~\ref{conformaldiagram}). The existence of an event 
horizon is determined by the convergence of the integral,
\begin{equation}
\eta_{\mathrm{max}} \equiv \int_{t_0}^{\infty} \frac{\mathrm{d}t}{a(t)} \label{etamax}, 
\end{equation}
which also implies that the conformal time is bounded in the future;
indeed the two conditions are equivalent. 
Just as we have defined the conformal time remaining
in Equation~\ref{etamax}, we are equally at liberty to determine
if the universe had a finite conformal past. The limits on the
Equation~\ref{etamax} would then be changed to integrate from $t =0$
to $t = t_0$. Finite values of either integral correspond
to the beginning, $\eta_\mathrm{min}$, and end, $\eta_\mathrm{max}$, of
the universe in conformal coordinates. 

The event horizon is distinct from the particle horizon, which is a
surface that divides all fundamental particles into two classes: 
those that have already been observable at the
present, $t_0$, and those that have not. [See \citet{Rindler} for
further details]. In other words, the particle
horizon is equal to the path of a photon originated from the Big Bang
(see the cyan curve in Figure~\ref{conformaldiagram}).
We already noticed that an event horizon only exists in universes with a finite
conformal future, likewise a particle horizon only exist 
in universes that have a finite conformal past.

If the universe has a flat spatial geometry and if it contains only
a single cosmic fluid with an equation of state $w \neq -1$, then 
$a(t) \propto t^n$, with $n = \frac{2}{3 (1 + w)}$. 
For a de Sitter universe ($\Omega_{\Lambda,0} = 1, \ \Omega_{m,0} = 0$),
$w=-1$, and we have the special case that $a(t) = \mathrm{e}^{H_0 t}$. 
From these expressions for the scale factor, we
can see that the integral in Equation~\ref{etamax} remains finite if
and only if $w < -1/3$. 
If we change the limits to integrate from $t =0$
to $t = t_0$, the integral would remain finite if
and only if $w > -1/3$. 
Thus single component flat universes with $w < -1/3$
have an event horizon only, while they have a particle
horizon only if $w > -1/3$.
If $w = -1/3$, then such a universe
neither has an event horizon, nor a particle horizon.
Because our universe was previously dominated by radiation
($w_\mathrm{r} = + 1/3$) and matter ($w_\mathrm{m} = 0$), it has a
particle horizon, and it also has an event horizon because it is
currently dominated by dark energy, which must have an equation of
state $w < -1/3$ for cosmic acceleration.

\subsection{The Hubble sphere}
The Hubble sphere marks the surface at which comoving systems are
receding from an observer at the speed of light according to Hubble's
law 
\begin{equation}
v_{\mathrm{rec}} \equiv H R, \label{hubble}
\end{equation} 
that is an object sitting on the Hubble sphere would have a recessional
velocity, $v_{\mathrm{rec}}=c$~\citep{Harrison1991, Davis}. 
Any object more distant than the Hubble sphere
is receding from us at a speed greater than the speed of light. An
object at a distance $R$ away has two components to its velocity,
which may be written as in terms of a recessional and peculiar
velocity as follows,
\begin{equation}
\dot{R} = \dot{a}r + a\dot{r} = v_{\mathrm{rec}} + v_{\mathrm{pec}}.
\end{equation} 
It is important to distinguish between these velocities; although the
recessional velocity may be greater than $c$, locally the peculiar
velocity is always subluminal. In fact, a greater than light speed
velocity is only inferred from non-local comparisions; if the velocity
vectors were parallel propagated along a null geodesic and then a
measurement of the redshift was taken, the resultant velocity would be
less than the speed of light~\citep{bunnhogg}. Thus, we can already
see from the definition of the Hubble sphere that we must be careful 
drawing conclusions with regards to its physical meaning.

The ``cosmic horizon'', or characteristic radius at which
$R_\mathrm{h} = 1/H$ in~\cite{melia2007, melia2009} and MA09,
is nothing more than the boundary of the Hubble sphere,
the Hubble surface. Remembering that we have set $c = 1$,
this is immediately clear from Equation~\ref{hubble}.
It is well documented in the literature that the Hubble sphere does not
constitute a true horizon, nor are events outside the Hubble sphere
permanently hidden from the observer's view~\citep{Harrison1981,
Davis}. Although photons emitted toward the observer by objects inside
the Hubble sphere approach the observer, whereas those emitted by
galaxies outside the Hubble sphere recede, if the Hubble parameter $H$
decreases with time, $R_\mathrm{h}$ will increase and overtake light
rays which were initially beyond the ``cosmic horizon''. It is the
particle horizon, rather than the Hubble sphere that defines the
furtherest distance from which we can receive a signal at the present
time. In fact, for the concordance cosmology, the Hubble surface
currently lies at $z \approx 1.5$ \citep{Davis} and, as any extragalactic astronomer
will attest, is certainly not a limit to how far we can observe.

There are two exceptions, however, for which the Hubble sphere does
constitute a horizon that cannot be traversed.
In these cases, it is degenerate with the particle horizon or 
with the event horizon for every cosmic instant.
In other words, the slope of the Hubble surface in a conformal diagram
(the red line in Figure~\ref{conformaldiagram}) is always $\pm 1$,
because the slope of the particle horizon in a
conformal diagram is $+1$, and the slope of the event horizon is $-1$.
To express the ``cosmic horizon'' in terms of the
comoving coordinate $r$, we use Equation~\ref{Rar}. 
This gives $r_\mathrm{h} = R_\mathrm{h}/a$.  
The slope of the Hubble surface in
in a conformal diagram is therefore equal to
\begin{equation}
\frac{\mathrm{d}r_\mathrm{h}}{\mathrm{d}\eta} = 
a \dot{r}_\mathrm{h} = - \frac{\ddot{a}a}{\dot{a}^2} .
\end{equation}
and we arrive at the definition of $q$ given earlier in
Equation~\ref{decel}. Note that $q$ is only constant in single
component universes, thus the Hubble surface is not a
cosmological horizon at all, except when it becomes degenerate with
the particle horizon in universe with radiation only ($q = 1$) and
with the event horizon in a de Sitter universe ($q = -1$).
[See also \citet{Harrison1991}.]


\section{Redshifting in the observer- dependent form of the metric}\label{redshifting}

MA09 originally showed that for $\mathrm{d}R=\mathrm{d}\theta=\mathrm{d}\phi=0$,
the time interval $\mathrm{d}t$ in
the observer-dependent form of the metric (Equation~\ref{meliametric}) must go to
infinity as $R \to R_\mathrm{h}$, leading them to conclude that the
``cosmic horizon'' is like the event horizon of a black hole, infinitely
redshifting any emission coinciding with it. This in contrast to
\citet{Davis}, who point out that redshift does not go to infinity for
objects on our Hubble sphere (in general) and for many 
cosmological models we can see beyond it.  
Here, we examine observed redshift of a photon exchanged between two 
observers in the observer-dependent coordinates.
As is apparent in Figure~\ref{conformaldiagram}, this redshift
should not go to infinity.

It is straightforward to demonstrate that the 4-velocity of any comoving observer
(which has fixed spatial coordinates in the FLRW metric) is given by
\begin{equation}
u^\alpha = ( 1 , \dot{a} r , 0 , 0 )
\label{fourvelocity}
\end{equation}
Furthermore, from Killing's equation, it can be
demonstrated that this spacetime admits a Killing vector of the form
\begin{equation}
\xi^\alpha = ( 0 , a k(\theta,\phi) , l(\theta,\phi) , m(\theta,\phi) )
\label{killing}
\end{equation}
where $k$, $l$ and $m$ represent three currently undetermined functions; as
we will be considering photons paths in the $(t,R)$ plane their exact
form is unnecessary. The existence of the Killing vector allows to
define a quantity, $e$, which is conserved along the geodesic path of
the photon, namely
\begin{equation}
e = \bxi \cdot {\bf p} = \left( \frac{R}{R_\mathrm{h}}\right) a p^t - a p^R
\label{killingcon}
\end{equation}
where $p^t$ and $p^R$ are the components of the photon 4-momentum
and the function $k$ is subsumed into the constant~$e$.

If two observers exchange a photon, the energy of the photon as seen by the receiver, $E_\mathrm{r}$, compared
to the energy as measured by the emitter, $E_\mathrm{e}$, is simply given by
\begin{equation}
\frac{E_\mathrm{r}}{E_\mathrm{e}} = \frac{ -{\bf u_r} \cdot {\bf p}(\mathrm{r}) } { -{\bf u_e} \cdot {\bf p}(\mathrm{e}) }
\label{photonenergy}
\end{equation}
where the ${\bf u}$ are the 4-velocities of the receiver and the
emitter, while ${\bf p}$ is the 4-momentum of the photon. In general,
we would have to propagate the photon between the emitter and the
receiver, although the presence of the Killing vector allows us to
simplify this procedure by noting that
\begin{equation}
E = -{\bf u} \cdot {\bf p}  = -\left(\Phi  p^t +
 \left( \frac{R}{R_\mathrm{h}}\right) p^R   + \left(\left( \frac{R}{R_\mathrm{h}}\right)  p^t - p^R\right) \dot{a}r \right)
\label{killingcom}
\end{equation}
then
\begin{equation}
E = -\frac{e}{a} \left( \frac{R}{R_\mathrm{h}} - \frac{p^R}{p^t} \right)
\end{equation}

The ratios of the components of the photon 4-momentum can be
determined from the metric (Equation~\ref{meliametric}), remembering that
photons follow null paths $(\mathrm{d}s=0)$ and so
\begin{equation}
\frac{\mathrm{d}R}{\mathrm{d}t} = \frac{p^R}{p^t} = \frac{ - \Phi}{\big(\frac{R}{R_\mathrm{h}}\big) \pm 1 }
\end{equation}
Following a photon from a positive $R$ to the origin selects the solution that
\begin{equation}
\frac{p^R}{p^t}  = - \left( 1 - \frac{R}{R_\mathrm{h}} \right)
\end{equation}
and clearly
\begin{equation}
E = -\frac{e}{a}
\end{equation}
Given this, a photon exchanged between two observers on the observer-dependent form of the FLRW
metric (Equation~\ref{meliametric}) will be see to have an energy dependent upon the
scale factor, $a$, at the time of emission and receipt, such that
\begin{equation}
\frac{E_\mathrm{r}}{E_\mathrm{e}} = \frac{a_\mathrm{e}}{a_\mathrm{r}} = \frac{1}{1+z}
\end{equation}
precisely the form seen in comoving coordinates (as expected). 

\begin{figure} 
\begin{center}
\includegraphics[scale = 0.45,angle=270]{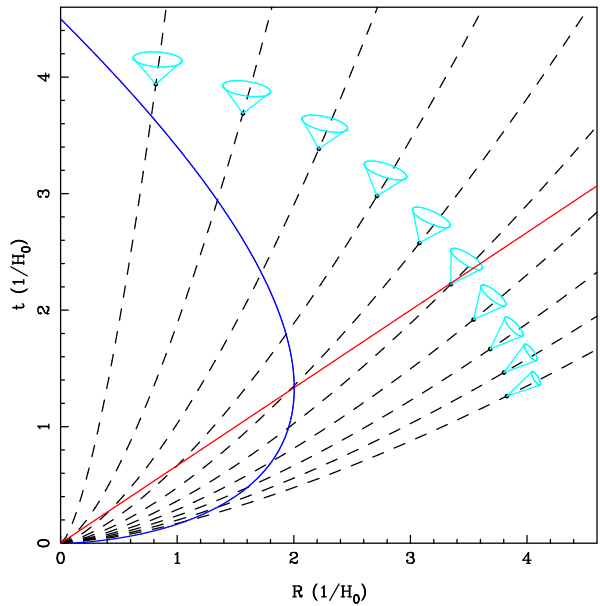}
\caption{Spacetime diagram in observer-dependent coordinates $(t,R)$
for a Einstein-de Sitter ($w=0$) universe, using the
metric in Equation~\ref{meliametric}. The Hubble sphere or cosmic
horizon is given by the red curve, while the blue line represents
a lightcone. Dashed lines represent the paths of comoving observers with their
lightcones in cyan; although stationary in $r$, their proper distance
$R$ increases. }
\label{mplt}
\end{center}
\end{figure} 

Figure~\ref{mplt} shows a spacetime diagram in the observer-dependent coordinates
used by MA09 for a universe containing a single component with
$w=0$. The blue line is a past lightcone at the moment the universe
is about 4.5 Hubble times old. The red line is the
``cosmic horizon'' and the dashed lines are worldlines from comoving
observers. As seen in this figure, photon paths (a past lightcone)
can extend through the ``cosmic horizon'' and hence objects even on the
``cosmic horizon'' are seen with a finite redshift $z$.
The shape of the lightcone would be exactly the same 
at any other moment in time, as would be the behaviour of the
$R_\mathrm{h}$ for other values of $w>-1$. 

It is interesting to note that, in examining the past lightcone in
Figure~\ref{mplt}, the ``cosmic horizon'' marks the turn-around point
for a photon path, a transition between a the photon moving away and
then moving towards us, and hence our past lightcone only encompasses
events with $R \leq R_\mathrm{h}$, although the emission from an
object on the horizon is not infinitely redshifted. We return to this
point in the next section.

\subsection{Metric Divergence}

As was shown in the previous section, $R_\mathrm{h}$ corresponds to a
stationary point in the past lightcone, where the trajectory changes
from moving away from the big bang to moving towards us.  The analysis
of MA09 considers the path of objects with {\bf fixed} R,
such that $u^R=0$; what do these correspond to?  By examining the
lightcone structure as we approach the ``cosmic horizon'', it is
apparent that such a trajectory is approaching the left-hand side of
the lightcone, implying that compared to a comoving observer at that
point they are moving closer and closer to the speed of light.
Remembering that for our comoving observer, $u^t = dt/d\tau = 1$
(where $\tau$ is the proper time registered by the comoving observer)
and for the fixed observer of MA09 $u^t=\Phi^{-\frac{1}{2}}$, 
it is apparent that the divergence is time
dilation between the comoving and the fixed observer, going to
infinity at $R_\mathrm{h}$ where the fixed observer is forced to
travel at the speed of light.  In summary, the divergence noted by
MA09 is due to forcing unphysical properties on the emitter
by requiring $\mathrm{d}R=0$.
If these unphysical properties were not demanded, the 
observer-dependent coordinates could be used to describe the spacetime
geometry, as long as one neglects the singularity
in the metric as $R \to R_\mathrm{h}$.


\section{Conclusion}

The inferred cosmic acceleration presents a conceptual dilemma; there
is abundant observational evidence that favours the existence of a
cosmological constant, yet some predictions and consequences of
$\Lambda$CDM remain so puzzling that modern cosmology is littered with
alternate mechanisms for reproducing the observational signatures of
accelerated expansion. The similarity between the $1/H_0$ and
the current age of the universe as pointed out by~MA09 and \citet{Lima},
as well as the original coincidence that $\Omega_{\Lambda,0} \sim
\Omega_{m,0}$, are genuinely problematic. While it is surprising that
the average deceleration parameter should be close to zero at this particular
instant in cosmic time, and may signify that aspects of the standard
model are contrived, arguments of this nature can not be prioritised
over constraints from observations. The near equality of the Hubble
surface $R_\mathrm{h}$ and the age of the universe $t_0$ requires a
cautious interpretation and does not immediately exclude a
cosmological model with a non-zero cosmological constant.
Furthermore, it is worth emphasising that a single spacetime geometry
may be expressed in several different coordinate systems and not all
features of the metric necessarily contain a physical meaning; a
poignant example is provided by the divergence at the event horizon in
Schwarzschild metric, which may be shown to be a coordinate
singularity when written in Eddington-Finkelstein
coordinates. MA09 used observer-dependent
coordinates to argue that $R_\mathrm{h}$ is a true horizon,
while the theoretical framework to infer any
conclusions about cosmic horizons was already there.
Since $R_\mathrm{h}$ is the Hubble surface,
it is not a physical horizon at which an infinite redshift is measured
(except in a de Sitter universe), despite the divergence in the 
observer-dependent form of the FLRW metric.

\section*{Acknowledgments}
PvO thanks the University of Sydney for hosting him
during his Masters research.
We acknowledge support from  ARC Discovery Project  DP0665574.

\end{document}